\documentclass[11pt]{article}
\usepackage{amssymb,amsmath,epsfig}
\def\one{{{{\rm 1} \kern -.19em {\rm l}}}}
\def\C{{{{\rm {\mbox{\small l}}} \kern -.50em {\rm C}}}}
\def\R{{{{\rm l} \kern -.15em {\rm R}}}}
\def\N{{{{\rm l} \kern -.15em {\rm N}}}}
\def\E{{{{\rm l} \kern -.15em {\rm E}}}}
\def\P{{{{\rm l} \kern -.15em {\rm P}}}}
\def\Z{{{{\rm Z} \kern -.35em {\rm Z}}}}
\def\1{{{{\rm 1} \kern -.35em {\rm 1}}}}

\begin{document}
\begin{sloppypar}
\vspace*{0cm}
\begin{center}
{\setlength{\baselineskip}{1.0cm}{ {\Large{\bf
EXACTLY-SOLVABLE QUANTUM SYSTEMS IN TERMS OF LAMBERT-W FUNCTIONS
\\}} }}
\vspace*{1.0cm}
{\large{\sc{Axel Schulze-Halberg$^\ast$}} and {\sc{Artur M. Ishkhanyan$^{\dagger \ddagger}$}}}
\end{center}
\noindent \\
$\ast$ Department of Mathematics and Actuarial Science and Department of Physics, Indiana University Northwest, 3400 Broadway,
Gary IN 46408, USA, E-mail: axgeschu@iun.edu \\ \\
$\dagger$ Russian-Armenian University, Yerevan, 0051 Armenia \\ \\
$\ddagger$ Institute for Physical Research, NAS of Armenia, Ashtarak, 0203 Armenia
\vspace*{.5cm}
\begin{abstract}
\noindent
We construct a variety of new exactly-solvable quantum systems, the potentials of which are given in terms of
Lambert-W functions. In particular, we generate Schr\"odinger models with energy-dependent potentials,
conventional Schr\"odinger models using the supersymmetry formalism, and two-dimensional Dirac systems.
In addition, we derive Wronskian integral formulas for Lambert-W functions.
\end{abstract}


\noindent \\

\section{Introduction}
The Lambert-W function, also known as product logarithm, is defined as a branch of the multivalued inverse
to the mapping $w \mapsto w \exp(w)$. While this function was first studied in \cite{lambert} \cite{euler}, its importance
and usefulness became clear only recently. In fact, the Lambert-W function appears in a wide variety of applications
throughout the sciences, including photovoltaic cell models \cite{xu}, hydraulic systems \cite{brkic},
ecological and evolutionary problems \cite{lehtonen}, and superradiance phenomena \cite{puente}. For
a detailed discussion of the Lambert-W function, its properties, and its applications, the reader may refer to
\cite{corless}. Besides the aforementioned examples, the Lambert-W function was recently found to
play an important role in two cases of exactly-solvable Schr\"odinger equations. The first case is associated
with an attractive potential that is singular at the origin \cite{artur}, and the second case involves a
step-like potential \cite{artur2}. The general solutions of the governing Schr\"odinger equations can be given
in terms of confluent hypergeometric functions. It is remarkable that even though the first potential admits
solutions of bound-state type, the confluent hypergeometric functions contained in those bound-state wave-functions do not degenerate to polynomials. Let us also mention that the above results on Schr\"odinger systems were generalized
to the Klein-Gordon equation \cite{tarloyan}. The purpose of the present work is to generalize the results presented
in \cite{artur} \cite{artur2} \cite{tarloyan}. In particular, we will use a variety of techniques in order to generate exactly-solvable Schr\"odinger and
Dirac models that feature potentials in terms of the Lambert-W function. The first type of potentials we
will construct here pertain to the Schr\"odinger equation and  are energy-dependent. Systems with energy-dependent
potentials are used for modeling e.g. magneto-hydrodynamic models of the dynamo
effect \cite{dynamo}, the Hamiltonian formulation of relativistic quantum mechanics \cite{rela2}, and
confined models \cite{lombard}. Models with energy-dependent potentials require the use of a modified
quantum theory that affects norm and completeness relation in a fundamental way \cite{formanek}. The second type of
potentials admitting exactly-solvable Schr\"odinger equations are generated by means of the
quantum-mechanical supersymmetry formalism (SUSY), sometimes referred to as Darboux transformation. This
formalism consists of several algorithms, a description and applications of which can be found in e.g.
\cite{cooper} \cite{djsusy} \cite{xbatconfluent} and references therein. As a byproduct of the SUSY formalism, we
obtain integral formulas for Lambert-W functions in terms of Wronskians \cite{xbatdiff}. The third and last type of
potential we derive here pertains to the massless Dirac equation in two dimensions that is
governing charge carrier transport in Dirac materials \cite{cayssol}.
The remainder of this article is organized as follows. While the preliminaries section 2
summarizes basic theory needed in this note, section 3 is devoted to the construction of energy-dependent potentials that are expressed in terms of the Lambert-W function.
Section 4 focuses on the generation of SUSY partners to the singular Lambert-W potential studied in \cite{artur},
and in section 5 matrix potentials expressed through Lambert-W functions for the massless Dirac equation are
constructed.

\section{Preliminaries}
In order to be self-contained, in this section we give brief reviews of the SUSY formalism, its algorithms and
associated integral formulas. Furthermore, we state the boundary-value problem for the Schr\"odinger equation with
singular Lambert-W potential \cite{artur}, along with its general solution.

\subsection{The SUSY formalism}
The algorithms of the quantum-mechanical SUSY formalism depart from the stationary Schr\"odinger equation
that can be written in the form
\begin{eqnarray}
\psi''(x)+\left[E-V_1(x) \right]~\psi(x) &=& 0. \label{susy1}
\end{eqnarray}
As usual, the constant $E$, referred to as energy, is an arbitrary constant and the potential $V$ is a continuous,
real-valued function. We must now distinguish the standard and
the confluent SUSY algorithms.
\paragraph{The standard SUSY algorithm.}
For a natural number $n$, assume that $u_1,u_2,...,u_n$ are solutions to equation (\ref{susy1}), associated to the
pairwise different energies $\lambda_1, \lambda_2,...,\lambda_n$, respectively. While the latter functions are
often referred to as transformation functions or auxiliary solutions, their associated energies are usually called
factorization energies. Now, the function
\begin{eqnarray}
\phi(x) &=& \frac{W_{u_1,u_2,...,u_n,\psi}(x)}{W_{u_1,u_2,...,u_n}(x)}, \label{susy2}
\end{eqnarray}
where each $W$ stands for the Wronskian of the functions in its index, is a solution to the transformed equation
\begin{eqnarray}
\phi''(x)+\left[E-V_2(x) \right]~\phi(x) &=& 0, \label{susy3}
\end{eqnarray}
the potential $V_2$ of which is related to its initial counterpart $V_1$ as
\begin{eqnarray}
V_2(x) &=& V_1(x) -2~\frac{d^2}{dx^2}~ \log\left[W_{u_1,u_2,...,u_n}(x) \right]. \label{susy4}
\end{eqnarray}
The function (\ref{susy2}) is called a SUSY transformation of order $n$. Note that the family
$(u_1,u_2,...,u_n,\psi)$ must be linearly independent in order to avoid that (\ref{susy2}) vanishes. If (\ref{susy1}) is the
governing equation of a spectral problem for the spectral parameter $E$, then a SUSY transformation of order $n$ can
change the spectrum of the transformed problem by at most $n$ values, depending on the factorization energies
$\lambda_1, \lambda_2,...,\lambda_n$. We will pick this topic up when stating various examples in section 4.
For a general discussion the reader is referred to the literature mentioned at the beginning of this paragraph.

\paragraph{The confluent SUSY algorithm and integral formulas.}
In the confluent version of the SUSY formalism, all factorization energies are equal. More precisely,
for a natural number $n \geq 2$, assume this time that the family of transformation functions $(u_1,u_2,...,u_n)$ is
a solution of the following system
\begin{eqnarray}
u_1''(x)+\left[\lambda-V_1(x) \right]~u_1(x) &=& 0 \label{j1} \\[1ex]
u_j''(x)+\left[\lambda-V_1(x)\right]~u_j(x) &=& -u_{j-1}(x),~~~j=2,...,n. \label{j2}
\end{eqnarray}
where the factorization energy $\lambda$ is an arbitrary constant. While we will regard the transformation functions
as depending on the variables $x$ and $\lambda$, for the sake of simplicity we do not include $\lambda$ as an
argument. Since $u_1,...,u_n$ can be identified as
generalized eigenvectors of the Hamiltonian associated with equation (\ref{j1}), the set $(u_1,u_2,...,u_n)$ is called
Jordan chain of order $n$. Now, if we assume that (\ref{j1}), (\ref{j2}) hold, then the function (\ref{susy2}) is a solution to
equation (\ref{susy3}) for the
potential (\ref{susy4}). It is important to point out that despite (\ref{susy2})-(\ref{susy4}) having the same form for the
standard and the confluent algorithm, they in general deliver different results due to the inequivalent constraints on the
transformation functions $u_1,u_2,...,u_n$ in each case. The general solution to the system (\ref{j1}), (\ref{j2}) can be
represented through integrals or derivatives \cite{xbatconfluent}. In the latter case, which we focus on in this note,
we assume that $v_1$ and $v_2$ are linearly
independent solutions of equation (\ref{j1}). Then, for $n \geq 2$ we have \cite{bermudezthesis}
\begin{eqnarray}
u_j(x) &=& \sum\limits_{j=1}^{n-1} \frac{C_{n-j}}{(j-1)!}~\frac{\partial^{j-1} ~v_1(x)}{\partial \lambda^{j-1}} +
\frac{D_{n-j}}{(j-1)!}~\frac{\partial^{j-1} v_2(x)}{\partial \lambda^{j-1}}+\frac{1}{(n-1)!}
~\frac{\partial^{n-1} v_1(x)}{\partial \lambda^{n-1}}. \label{difform}
\end{eqnarray}
Here, we applied partial derivatives with respect to $\lambda$, recall that we assumed the transformation functions
to depend on both $x$ and $\lambda$, but omit to include the latter variable as an argument. In (\ref{difform})
we introduced arbitrary constants $C_j,~ D_j$, $1 \leq j \leq n-1$, note that these constants are allowed to be
zero. In the particular case $n=2$, our identity (\ref{difform}) simplifies to
\begin{eqnarray}
u_2(x) &=& C_0~v_1(x)+D_0~v_2(x)+\frac{\partial v_1(x)}{\partial \lambda}. \label{diff2}
\end{eqnarray}
Before we conclude this section, let us mention
an application of the confluent SUSY formalism to integration. Assuming that $u_1$ is a solution of (\ref{j1}), we have
the following identities for single and double integration
\begin{eqnarray}
\int\limits_{p}^x u_1(t)^2~dt &=& W_{u_1,\frac{\partial u_1}{\partial \lambda}}(p)-
W_{u_1,\frac{\partial u_1}{\partial \lambda}}(x) \label{int1} \\[1ex]
\int\limits_{p}^x \int\limits_{p}^{x_1} \left[\frac{u_1(x_2)}{u_1(x_1)}\right]^2 dx_2~dx_1 &=&
\frac{{\displaystyle{\frac{\partial}{\partial \lambda}~u_1(p)}}}{u_1(p)}-
\frac{{\displaystyle{\frac{\partial}{\partial \lambda}~u_1(x) }}}{u_1(x)}+
W_{u_1,\frac{\partial}{\partial \lambda}u_1}(p)~
\int\limits_{p}^x \frac{1}{u_1(t)^2}~dt, \label{int2}
\end{eqnarray}
where $p$ is any value inside the domain of $u_1$. It is important to point out that the identities (\ref{int1}) and
(\ref{int2}) only hold if the integrals exist. The advantage of those identities consists in
the fact that integration can be performed entirely by finding partial derivatives. For generalizations to higher-order
integrals and further discussion the reader is referred to \cite{xbatdiff}.

\subsection{The singular Lambert-W model}
We summarize results from \cite{artur}. The system we will focus on in this work is governed by the following boundary-value problem of Dirichlet-type
\begin{eqnarray}
\psi''(x) + \left\{E - V_0+\frac{V_0}{1+W \hspace{-.1cm}\left[
-\exp\left(
\frac{x_0-x}{\sigma}
\right)
\right]
}
\right\} \psi(x) &=& 0,~~~x \in (\sigma+x_0,\infty) \label{bvp1}
\\[1ex]
\psi(\sigma+x_0) ~=~ \lim\limits_{x \rightarrow \infty} \psi(x) &=& 0, \label{bvp2}
\end{eqnarray}
where $E$, $\sigma>0$, $V_0>0$ and $x_0$ are real-valued constants. Furthermore, $W$ stands for the product
logarithm or Lambert-W function. Since this function is in general complex and multivalued,
the left endpoint $x=\sigma+x_0$ of the problem's boundary is determined by
the requirement that the Lambert-W function take unique and real values. Our Schr\"odinger equation (\ref{bvp1})
features the interaction
\begin{eqnarray}
V(x) &=& V_0-\frac{V_0}{1+W \hspace{-.1cm}\left[
-\exp\left(
\frac{x_0-x}{\sigma}
\right)
\right]
}, \label{pot}
\end{eqnarray}
which we will refer to as singular Lambert-W potential. The singularity of our interaction (\ref{pot}) is located at
the left domain endpoint $x=\sigma+x_0$. The boundary-value problem (\ref{bvp1}), (\ref{bvp2}) admits
a discrete energy spectrum $(E_n)$, the values of which are determined as solutions of the transcendental equation
\begin{eqnarray}
1+\frac{(s-c)~{}_1F_1\left(a+1,c+1,s \right)}{2~c~ {}_1F_1\left(a,c,s \right)} &=& 0, \label{trans}
\end{eqnarray}
with respect to $E$, where the following abbreviations are in use
\begin{eqnarray}
a ~=~ -\frac{\left(\sqrt{-E}-\sqrt{V_0-E} \right)^2\sigma}{2~\sqrt{V_0-E}} \qquad \qquad
c ~=~ 2~\sqrt{-E}~\sigma \qquad \qquad
s ~=~ 2~\sqrt{V_0-E}~\sigma. \label{acs}
\end{eqnarray}
Equation (\ref{trans}) has a finite number of $N$ solutions $E_n$, $0 \leq n \leq N-1$, that increases as the parameters
$V_0$ and $\sigma$ are raised. If the values of the latter parameters are too small, (\ref{trans}) has no solutions.
For each spectral value $E_n$, $n = 0,1,2,...,N-1$, the boundary-value problem (\ref{bvp1}), (\ref{bvp2}) admits an
associated solution $\psi_n \in L^2(\sigma+x_0,\infty)$, the explicit form of which is given by
\begin{eqnarray}
\psi_n(x) &=& \exp\left\{\frac{c}{2}~W \hspace{-.1cm}\left[
-\exp\left(
\frac{x_0-x}{\sigma}
\right)
\right]
\right\}
W \hspace{-.1cm}\left[
-\exp\left(
\frac{x_0-x}{\sigma}
\right)
\right]^\frac{c}{2} \times \nonumber \\[1ex]
&\times&
\left\{\frac{d}{dz}~ \exp\left[
\frac{(c-s) z}{2}
\right] {}_1F_1(a,c,s~z) \right\}_{\mid z = -W \hspace{-.1cm}\left[
-\exp\left(
\frac{x_0-x}{\sigma}
\right)
\right]}, \label{solder}
\end{eqnarray}
where the symbol ${}_1F_1$ stands for the confluent hypergeometric function \cite{abram}.
Note further that we employed the abbreviations defined in (\ref{acs}), where now $E$ must be replaced by $E_n$.
After evaluation of the derivative in (\ref{solder}), the
solution reads
\begin{eqnarray}
\psi_n(x) &=& \exp\left\{\frac{c}{2}~W \hspace{-.1cm}\left[
-\exp\left(
\frac{x_0-x}{\sigma}
\right)
\right]
\right\}
W \hspace{-.1cm}\left[
-\exp\left(
\frac{x_0-x}{\sigma}
\right)
\right]^\frac{c}{2} \times \nonumber \\[1ex]
&\times& \Bigg(~ \frac{c-s}{2}~{}_1F_1\hspace{-.1cm} \left\{
a,c,-s~W \hspace{-.1cm}\left[
-\exp\left(
\frac{x_0-x}{\sigma}
\right)
\right]
\right\}+ \nonumber \\[1ex]
&+&\frac{a~s}{c}~{}_1F_1 \hspace{-.1cm}\left\{
a+1,c+1,-s~W \hspace{-.1cm}\left[
-\exp\left(
\frac{x_0-x}{\sigma}
\right)
\right]
\right\} \Bigg). \label{solfull}
\end{eqnarray}
The hypergeometric functions on the right side do not degenerate to polynomials because their first argument does not
take nonnegative integer values, as inspection of (\ref{acs}) shows.

\section{Energy-dependent Lambert-W potentials}
\begin{figure}[h]
\begin{center}
\epsfig{file=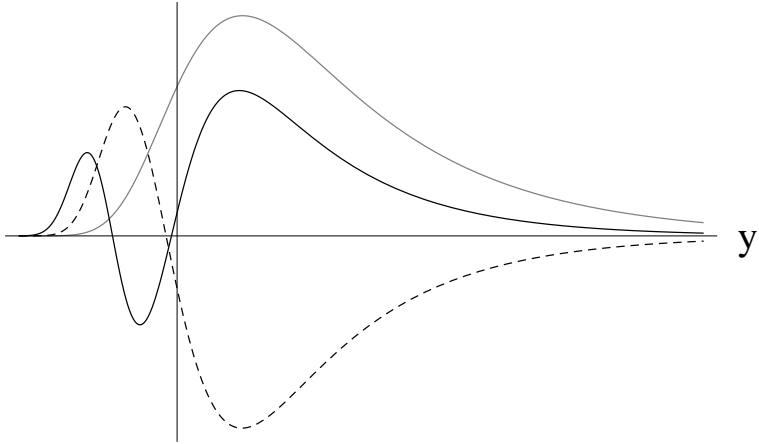,width=10cm}
\caption{Graphs of the functions $\phi_0$ (gray curve), $\phi_1$ (dashed curve) and $\phi_2$ (black curve)
for the parameter settings $\sigma=-x_0=V_0=5$.
The corresponding spectral values are approximately ${\cal E}_0 = 1.6539 \times 10^{-6}$,
${\cal E}_1 = 5.9717 \times 10^{-4}$,
${\cal E}_2 = 1.8975 \times 10^{-2}$. For better visibility the plots were scaled both
horizontally and vertically.
}
\label{e_sol}
\end{center}
\end{figure}

We will now demonstrate how generalizations of the model (\ref{bvp1}), (\ref{bvp2}) can be constructed that
feature energy-dependent potentials. The underlying idea is to apply a point transformation (coordinate change) to
the governing equation (\ref{bvp1}) of our initial boundary-value problem. Let us directly employ a particular
point transformation of the form
\begin{eqnarray}
x(y) ~=~\sigma+x_0+ \exp\left[-\frac{y}{(E-1)^4} \right] \qquad \qquad \qquad \phi(y) ~=~ \sqrt{\frac{1}{x'(y)}}~\psi[x(y)]. \label{pct}
\end{eqnarray}
Upon substitution of this transformation, our boundary-value problem (\ref{bvp1}), (\ref{bvp2}) is converted to the form
\begin{eqnarray}
\phi''(y) \hspace{-.2cm} &+& \hspace{-.2cm}  \left\{
~\frac{1}{2~(E-1)^8}-
\frac{3}{4~(E-1)^8~
\Bigg(
1+
W\left\{
-\exp\left[
-\frac{x(y)-x_0}{\sigma}
\right]
\right\}
\Bigg)}~- \right. \nonumber \\[1ex]
& & \left. \hspace{.3cm}
-~\frac{3~W\left\{
-\exp\left[
-\frac{x(y)-x_0}{\sigma}
\right]
\right\}
}
{4~(E-1)^8~
\Bigg(
1+
W\left\{
-\exp\left[
-\frac{x(y)-x_0}{\sigma}
\right]
\right\}
\Bigg)}~+ \right. \nonumber \\[1ex]
& & \left. \hspace{.3cm}
+~\frac{E+(E-V_0)~W\left\{
-\exp\left[
-\frac{x(y)-x_0}{\sigma}
\right]
\right\}
}{x(y)^2~(E-1)^8~
\Bigg(
1+
W\left\{
-\exp\left[
-\frac{x(y)-x_0}{\sigma}
\right]
\right\}
\Bigg)}
~\right\}~\phi(y) ~=~ 0,~~~y \in (-\infty,\infty) \nonumber \\ \label{bvp3} \\[1ex]
& & \hspace{5.85cm} \lim\limits_{y \rightarrow -\infty} \phi(y)~=~ \lim\limits_{y \rightarrow \infty} \phi(y) ~=~ 0, \label{bvp4}
\end{eqnarray}
where the function $x=x(y)$ represents the coordinate change that is defined in (\ref{pct}). We observe that
(\ref{bvp3}) can be interpreted as a Schr\"odinger equation for an energy-dependent potential.
It is important to note that the system's energy is not given by the constant $E$ anymore, but
by the constant term
that the coefficient of $\phi$ in (\ref{bvp3}) contains. We introduce the energy ${\cal E}$ of the model (\ref{bvp3}),
(\ref{bvp4}) as follows
\begin{eqnarray}
{\cal E} &=& \frac{1}{2~(E-1)^8}. \label{enee}
\end{eqnarray}
In order to obtain an explicit form of the actual energy-dependent potential contained in (\ref{bvp1}), we first
express the parameter $E$ through the energy ${\cal E}$. This gives
\begin{eqnarray}
E &=& 1-\left(
\frac{1}{2~{\cal E}}
\right)^\frac{1}{8}. \label{einv}
\end{eqnarray}
We remark that inversion of (\ref{enee}) is possible because $E$ is real-valued, such that ${\cal E}$ is strictly
monotone with respect to $E$. Upon substitution of (\ref{einv}) into (\ref{bvp1}), we can extract the energy-dependent
potential $U$ in the form
\begin{eqnarray}
U\left(y,{\cal E} \right) &=& -\frac{ \exp\left(
-2~\sqrt{2 ~{\cal E}}~ y
\right)
}{2~\left(
1+W\hspace{-.1cm}\left\{
-\exp\left[
-1-\frac{1}{\sigma}~ \exp\left(
-\sqrt{2 ~{\cal E}}~ y
\right)
\right]
\right\}
\right)} \times \nonumber \\[1ex]
&\times&
\Bigg( \hspace{-.1cm}
-2~(2~{\cal E})^\frac{7}{8}+\left[
4-3~\exp\left(
2~\sqrt{2 ~{\cal E}}~ y
\right)
\right] {\cal E}
+\left\{
-2~(2~{\cal E})^\frac{7}{8}
+ \right.    \nonumber \\[1ex]
&+& \left.
\left[
4-3~\exp\left(
2~\sqrt{2 ~{\cal E}}~ y
\right)-4~V_0
\right] {\cal E}
\right\}
W\hspace{-.1cm}\left\{
-\exp\left[
-1-\frac{1}{\sigma}~ \exp\left(
-\sqrt{2 ~{\cal E}}~ y
\right)
\right]
\right\}
\Bigg) \hspace{-.1cm}. \label{poteff}
\end{eqnarray}
\begin{figure}[h]
\begin{center}
\epsfig{file=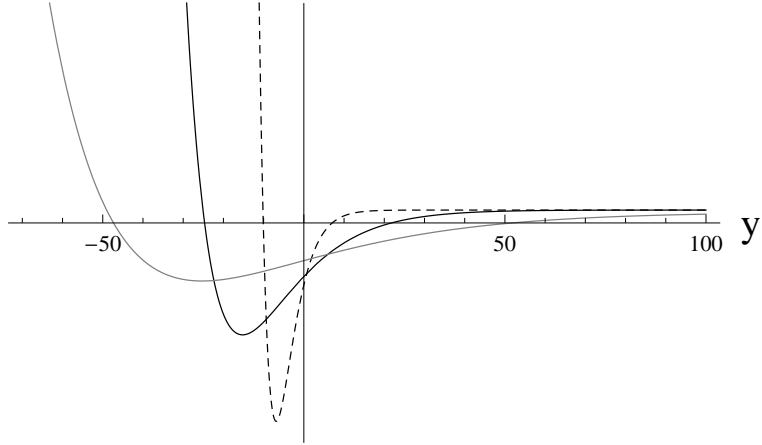,width=10cm}
\caption{Graphs of the energy-dependent potential (\ref{poteff}) for the parameter settings $\sigma=-x_0=V_0=5$ and
$E=3/2$ (gray curve), $E=-1$ (black curve), $E=-1/2$ (dashed curve). For better visibility the plots were scaled vertically.
}
\label{e_pot}
\end{center}
\end{figure} \noindent
Figure \ref{e_pot} shows graphs of the potential (\ref{poteff}) for different parameter settings. Note that by specifying a value
for $E$, the energy ${\cal E}$ is determined by means of (\ref{enee}).
If we combine the results (\ref{enee}) and (\ref{poteff}), we can write the governing equation (\ref{bvp3})
of our boundary-value problem as
\begin{eqnarray}
\phi''(y)+\left[{\cal E}-U\left(y,{\cal E} \right)\right] \phi(y) &=& 0. \nonumber
\end{eqnarray}
As mentioned before, this is a Schr\"odinger equation for the stationary energy ${\cal E}$ and the energy-dependent
potential $U$. Now, recall that the initial problem
(\ref{bvp1}), (\ref{bvp2}) has a finite discrete spectrum $(E_n)$. Therefore, any spectral value $E_n$ for $n \leq N$
determines a quantity ${\cal E}_n$ by means of our formula (\ref{enee}). Similarly, each bound-state solution
$\psi_n$, $n \leq N$, given in (\ref{solfull}), defines a solution $\phi_n$ of (\ref{bvp3}) by means of the point
transformation (\ref{pct}). In addition, standard estimates at the infinities show that the functions $\phi_n$ satisfy the
boundary conditions (\ref{bvp4}). Figure \ref{e_sol} visualizes these properties, showing graphs of the
solutions for a particular parameter setting.
 \noindent
Since we have now constructed solutions to the transformed
boundary-value problem (\ref{bvp3}), (\ref{bvp4}), it remains to verify their physical validity. To this end, we must
verify that the modified norm for systems featuring energy-dependent potentials exists and is positive.
In the present case, this norm is given by \cite{formanek}
\begin{eqnarray}
\Vert \phi_n \Vert &=& \int\limits_{-\infty}^\infty \left[1-\frac{\partial U(y,{\cal E})}{\partial {\cal E}} \right] |\phi_n(y)|^2~dy. \label{norm}
\end{eqnarray}
We remark that the integral on the right side of (\ref{norm}) is not a norm in the mathematical sense because it can
become negative. After substitution of the functions $\phi_n$ and
our potential in the form (\ref{poteff}), evaluation of the derivative and reinstalling the parameter $E$ by means of
(\ref{enee}) we obtain a very long and complicated expression that we omit to show here. Since we are not able to
integrate the latter expression symbolically, we restrict ourselves to calculating (\ref{norm}) for a particular
example by means of numerical integration. Upon choosing the parameter values $\sigma=-x_0=V_0=5$,
we obtain the following norms for the first three bound-state solutions
\begin{eqnarray}
\Vert \phi_0 \Vert ~=~  1.96965 \times 10^{-6} \qquad \qquad
\Vert \phi_1 \Vert ~=~  4.07894 \times 10^{-6} \qquad \qquad
\Vert \phi_2 \Vert ~=~  2.44908 \times 10^{-5}. \nonumber
\end{eqnarray}
Since these norms are positive, we conclude that the discrete spectrum (\ref{enee}) and its associated
bound-state solutions $\phi_n$, $0\leq n \leq N-1$, form a physically acceptable solution of the boundary-value
problem (\ref{bvp3}), (\ref{bvp4}).

\section{SUSY partners and integral formulas}
We will now calculate SUSY partners of the boundary-value problem (\ref{bvp1}), (\ref{bvp2}). To this end,
throughout this section we will refer to the initial potential (\ref{pot}) as $V_1$ instead of $V$. As a further
application, we work out the integral formulas (\ref{int1}) and (\ref{int2}) for particular cases.

\subsection{Conventional SUSY partners}

\begin{figure}[h]
\begin{center}
\epsfig{file=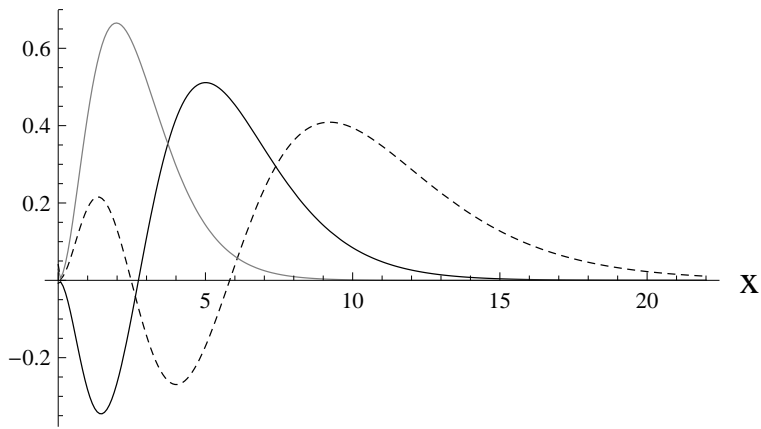,width=10cm}
\caption{Graphs of the SUSY-transformed and $L^2$-normalized functions $\phi_1$ (gray curve),
$\phi_2$ (black curve) and $\phi_3$ (dashed curve), as defined in (\ref{first}).
Parameter settings are $\sigma=-x_0=V_0=5$.}
\label{susyf}
\end{center}
\end{figure}

\begin{figure}[h]
\begin{center}
\epsfig{file=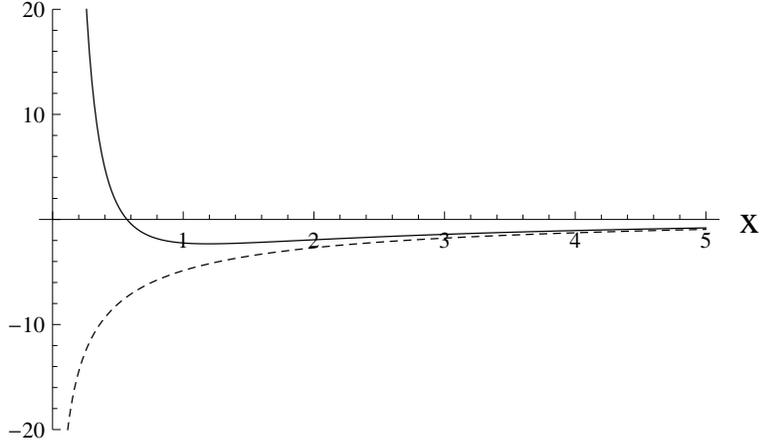,width=10cm}
\caption{Graphs of the SUSY-transformed potential (solid curve) and its initial counterpart (dashed curve), as given in
(\ref{tpot}) and (\ref{pot}), respectively. Parameter settings are $\sigma=-x_0=V_0=5$.}
\label{pot_susyf}
\end{center}
\end{figure}

\paragraph{First-order algorithm.}
In order to apply the SUSY formalism of first order, we need to determine a single transformation function $u_1$ that
is a solution to our governing equation (\ref{bvp1}). We choose the ground state function $\psi_0$ from the solution set
given in (\ref{solfull}), that is, we set
\begin{eqnarray}
u_1(x) &=& \psi_0(x). \label{u1}
\end{eqnarray}
This choice implies that the factorization energy $\lambda_1$ equals the ground state energy $E_0$. As a
consequence, the discrete spectrum of the transformed system will be missing the value $E_0$. According to
(\ref{susy2}) for $n=1$, the first-order SUSY transformation involving the function $u_1$ takes the form
\begin{eqnarray}
\phi_n(x) &=& -\frac{\psi_0'(x)}{\psi_0(x)}~\psi_n(x)+\psi_n'(x),~~~1 \leq n \leq N-1. \label{first}
\end{eqnarray}
Observe that the value $n=0$ was excluded, as it would lead to a vanishing function $\phi_0$. In principle we could
state the explicit form of (\ref{first}) by substituting (\ref{solfull}) and evaluating the derivatives.
However, since this would lead to a very long and complicated expression, we omit to specify the functions
(\ref{first}) any further. Figure \ref{susyf} shows these functions for a particular parameter setting.
 \noindent
They are solutions of the transformed boundary-value problem consisting of equation (\ref{susy3}), equipped with the
boundary conditions
\begin{eqnarray}
\phi_n(0) ~=~ \lim\limits_{x \rightarrow \infty} \phi_n(x) ~=~ 0,~~~1 \leq n \leq N-1. \nonumber
\end{eqnarray}
The transformed potential $V_2$ in equation (\ref{susy3}) is given by (\ref{susy4}) for $n=1$. Taking into account the
setting (\ref{u1}) and the form (\ref{pot}) of our initial potential $V_1$, we have
\begin{eqnarray}
V_2(x) &=& V_0-\frac{V_0}{1+W \hspace{-.1cm}\left[
-\exp\left(
\frac{x_0-x}{\sigma}
\right)
\right]
}-2~\frac{d^2}{dx^2}~ \log\left[\psi_0(x) \right], \label{tpot}
\end{eqnarray}
where $\psi_0$ is defined in (\ref{solfull}) for $n=0$. We do not show the explicit form of (\ref{tpot}) because the result
is a very long and complicated expression in terms of hypergeometric functions.
The transformed potential is visualized in figure \ref{pot_susyf},
along with its initial counterpart (\ref{pot}).
 \noindent

\begin{figure}[h]
\begin{center}
\epsfig{file=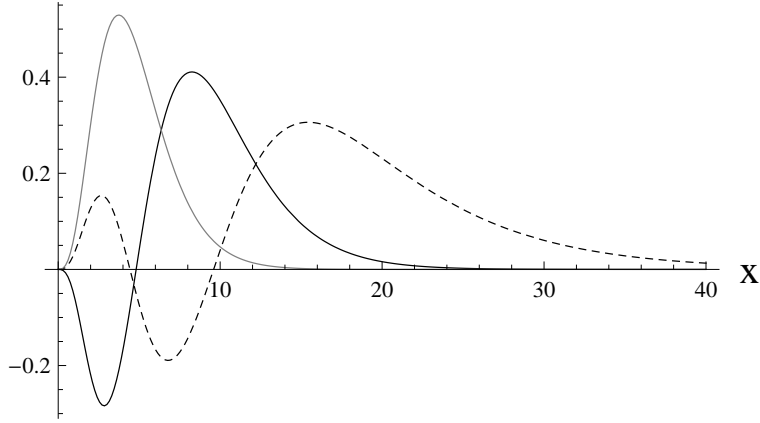,width=10cm}
\caption{Graphs of the SUSY-transformed and $L^2$-normalized functions $\phi_2$ (gray curve),
$\phi_3$ (black curve) and $\phi_4$ (dashed curve), as defined in (\ref{second}).
Parameter settings are $\sigma=-x_0=V_0=5$.}
\label{susys}
\end{center}
\end{figure}

\begin{figure}[h]
\begin{center}
\epsfig{file=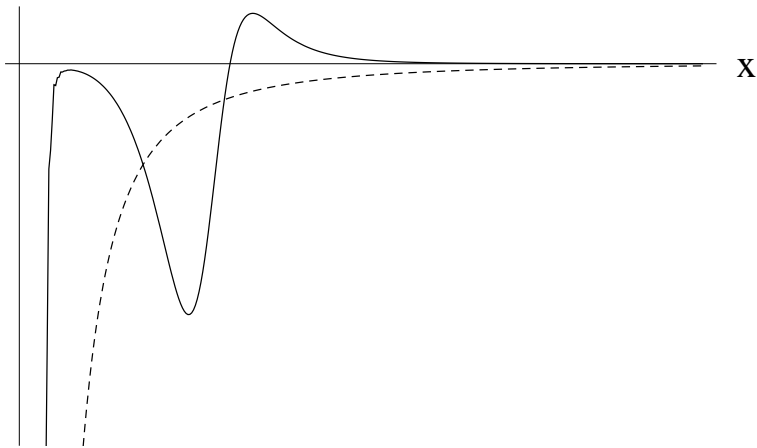,width=10cm}
\caption{Graphs of the SUSY-transformed potential (solid curve) and its initial counterpart (dashed curve), as given in
(\ref{tpot2}) and (\ref{pot}), respectively. Parameter settings are $\sigma=-x_0=V_0=5$. For better visibility, the plot is
scaled both vertically and horizontally.}
\label{pot_susys}
\end{center}
\end{figure}

\paragraph{Second-order algorithm.}
Performing a second-order SUSY transformation requires two transformation functions $u_1$ and $u_2$ that
solve our governing equation (\ref{bvp1}) at two different factorization energies $\lambda_1$ and $\lambda_2$,
respectively. Let us set
\begin{eqnarray}
u_1(x) ~=~ \psi_0(x) \qquad \qquad u_2(x) ~=~ \psi_1(x), \label{u12}
\end{eqnarray}
where the functions $\psi_1$ and $\psi_2$ were taken from (\ref{solfull}) for $n=1$ and $n=2$, respectively.
Our setting (\ref{u12}) implies that the values $E_0$ and $E_1$ will be erased from the discrete spectrum of the
SUSY-transformed system. We substitute $n=2$ into the general form (\ref{susy2}) of our SUSY transformation, which
gives after incorporation of (\ref{u12})
\begin{eqnarray}
\phi_n(x) &=& \frac{W_{\psi_0,\psi_1,\psi_n}(x)}{W_{\psi_0,\psi_1}(x)},~~~2 \leq n \leq N-1. \label{second}
\end{eqnarray}
We excluded the values $n=1$ and $n=2$ because they would give a vanishing SUSY transformation. We do not
plug the definition (\ref{solfull}) into our transformation (\ref{second}) because the resulting expression is not
manageable. Figure \ref{susys} visualizes (\ref{second}) for a particular parameter setting.
 \noindent
The functions $\phi_n$, $2 \leq n \leq N-1$, solve the boundary-value problem given by
equation (\ref{susy3}) for the boundary conditions
\begin{eqnarray}
\phi_n(0) ~=~ \lim\limits_{x \rightarrow \infty} \phi_n(x) ~=~ 0,~~~2 \leq n \leq N-1. \nonumber
\end{eqnarray}
The transformed potential $V_2$ in equation (\ref{susy3}) is given by (\ref{susy4}) for $n=2$. Taking into account the
setting (\ref{u12}) and the form (\ref{pot}) of our initial potential $V_1$, we have
\begin{eqnarray}
V_2(x) &=& V_0-\frac{V_0}{1+W \hspace{-.1cm}\left[
-\exp\left(
\frac{x_0-x}{\sigma}
\right)
\right]
}-2~\frac{d^2}{dx^2}~ \log\left[W_{\psi_0,\psi_1}(x) \right], \label{tpot2}
\end{eqnarray}
where $\psi_0$ and $\psi_1$ are defined in (\ref{solfull}) for $n=0$ and $n=1$, respectively.
As before, the explicit form of (\ref{tpot2}) for the present case is not shown due to its length. \
The transformed potential is shown in figure \ref{pot_susys},
along with its initial counterpart (\ref{pot}).
 \noindent

\subsection{Confluent SUSY partners}
\begin{figure}[h]
\begin{center}
\epsfig{file=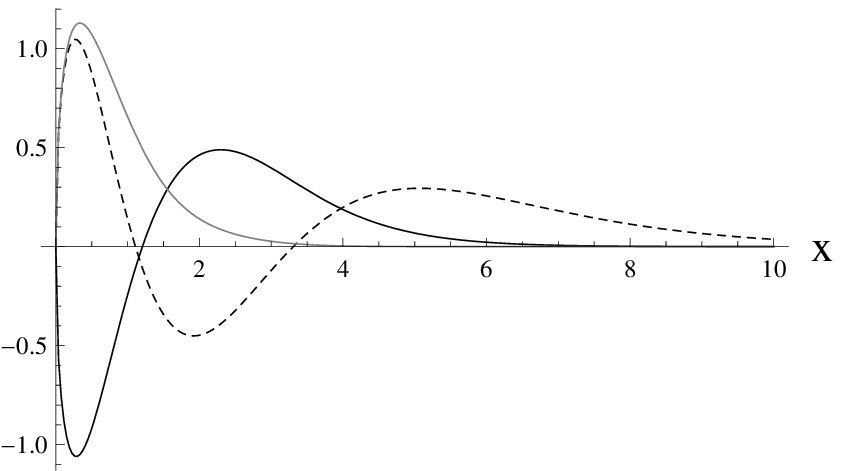,width=10cm}
\caption{Graphs of the SUSY-transformed and $L^2$-normalized functions $\phi_0$ (gray curve),
$\phi_1$ (black curve) and $\phi_2$ (dashed curve), as defined in (\ref{phicon}).
Parameter settings are $\sigma=-x_0=V_0=5$.}
\label{susyc}
\end{center}
\end{figure}

\begin{figure}[h]
\begin{center}
\epsfig{file=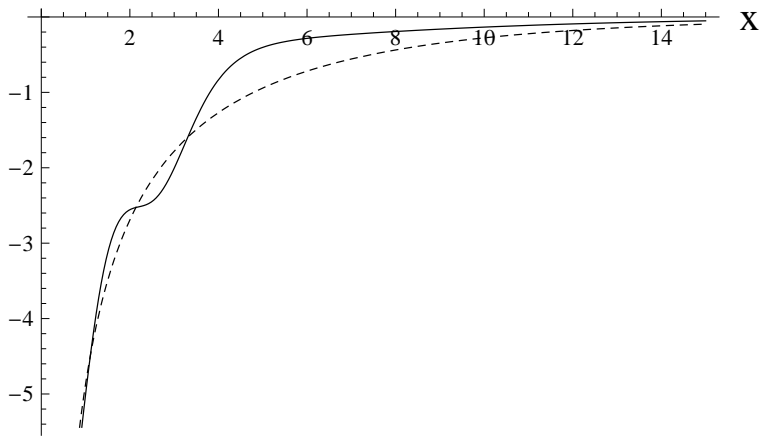,width=10cm}
\caption{Graphs of the SUSY-transformed potential (solid curve) and its initial counterpart (dashed curve), as given in
(\ref{tpotc}) and (\ref{pot}), respectively. Parameter settings are $\sigma=-x_0=V_0=5$.}
\label{potsusyc}
\end{center}
\end{figure}

The confluent SUSY algorithm in its simplest case requires two transformation functions $u_1$ and $u_2$, where
$u_1$ solves the initial Schr\"odinger equation (\ref{bvp1}) and $u_2$ is a solution to (\ref{j2}) for $j=2$. Since the
system of equations (\ref{j1}), (\ref{j2}) is coupled, the function $u_1$ can be generated from its counterpart $u_0$,
for example by means of the formula (\ref{diff2}). Choosing $C_0=D_0=0$ in the latter formula, we set
\begin{eqnarray}
u_1(x) ~=~ \psi_2(x) \qquad \qquad u_2(x) ~=~ \frac{\partial \psi_2(x)}{\partial E_2}, \label{u12c}
\end{eqnarray}
where $\psi_2$ is taken from (\ref{solfull}) for $n=2$. We remark that the simplified notation of the derivative in $u_2$
is understood as taking the derivative of the function $\psi_n$ for undetermined $E$ with respect to $E$, and afterwards
replacing $E$ by $E_2$. Note that in the confluent SUSY algorithm the settings (\ref{u12c})
do not imply a change in the discrete spectrum of the transformed system. Let us now
substitute $n=2$ into the form (\ref{susy2}) of the SUSY transformation and
incorporate our settings (\ref{u12c}). This gives
\begin{eqnarray}
\phi_n(x) &=& \frac{W_{\psi_2, \frac{\partial \psi_2}{\partial E_2},\psi_n}(x)}
{W_{\psi_2, \frac{\partial \psi_2}{\partial E_2}}(x)},~~~0 \leq n \leq N-1. \label{phicon}
\end{eqnarray}
As in the examples for the standard SUSY algorithm, the explicit form of the expressions contained in (\ref{phicon}) is
too large to be displayed here. Instead of showing these expressions, we visualize the functions (\ref{phicon}) for a
particular parameter setting, see figure \ref{susyc}.
 \noindent
According to the confluent SUSY algorithm, the functions in (\ref{phicon}) solve the boundary-value problem
consisting of the Schr\"odinger equation (\ref{susy3}) for the boundary conditions
\begin{eqnarray}
\phi_n(0) ~=~ \lim\limits_{x \rightarrow \infty} \phi_n(x) ~=~ 0,~~~0 \leq n \leq N-1. \label{bvpc}
\end{eqnarray}
Here, the transformed potential $V_2$ can be constructed by plugging $n=2$ into the general form (\ref{susy4}),
together with (\ref{u12c}) and (\ref{pot}). This yields
\begin{eqnarray}
V_2(x) &=& V_0-\frac{V_0}{1+W \hspace{-.1cm}\left[
-\exp\left(
\frac{x_0-x}{\sigma}
\right)
\right]
}-2~\frac{d^2}{dx^2}~ \log\left[W_{\psi_2, \frac{\partial \psi_2}{\partial E_2}}(x) \right], \label{tpotc}
\end{eqnarray}
note that $\psi_2$ and its derivative are obtained from (\ref{solfull}) for $n=2$. Instead of stating the general form
of the transformed potential, we show a special case of it in figure \ref{potsusyc}, along with the initial potential
(\ref{pot}).
 \noindent
Before we conclude this example, let us point out that the initial and SUSY-transformed system are isospectral, that is,
their discrete spectra coincide.

\subsection{Integral formulas}
Recall that our formulas (\ref{int1}) and (\ref{int2}) become applicable, if the function $u_1$ solves an equation
of the form (\ref{j1}). For the sake of simplicity, throughout this paragraph we will rename the function $u_1$ to
$u$.

\paragraph{Single integration.} We focus on the integral formula (\ref{int1}). As our function $u$ we choose the
solutions (\ref{solfull}) to the initial equation (\ref{bvp1}), where we assume that the energy $E$ does not equal
the spectral values $E_n$, but is kept arbitrary. Taking into account that $E$ plays the role of our parameter
$\lambda$ in (\ref{int1}), our integral formula reads
\begin{eqnarray}
\int\limits_{p}^x \psi(t)^2~dt &=& W_{\psi,\frac{\partial}{\partial E} \psi}(p) - W_{\psi,\frac{\partial}{\partial E} \psi}(x).
\label{intpsi}
\end{eqnarray}
Observe that we have removed the index from $\psi$ for denoting that we are not referring to the solutions of the full
boundary-value problem (\ref{bvp1}), (\ref{bvp2}), but only to the governing equation (\ref{bvp1}). As such,
the integral on the left side of (\ref{intpsi}) is not guaranteed to exist or be finite, this depends on the
particular choice of the parameters contained in $\psi$. While in general
our identity (\ref{intpsi}) does not have any physical meaning, its left side becomes an $L^2$-normalization
integral if $\psi$ is replaced by a bound-state solution $\psi_n$ from (\ref{solfull}) and the integration limits are
set to negative and positive infinity, respectively. In this case, the right side
of (\ref{intpsi}) must be evaluated, after which the parameter $E$ is replaced by the corresponding spectral
value $E_n$ emerging from (\ref{trans}). Due to the excessive length of the Wronskian in (\ref{intpsi}), we
do not state its explicit form, but instead show results for a particular parameter setting in table \ref{table}.
However, it is important to note that the right side of (\ref{intpsi}) provides a closed-form representation of the
indefinite integral (equal to a definite integral with undetermined limits of integration).
\renewcommand{\arraystretch}{2.2}
\begin{table}[h]
\begin{center}
\vspace{.5cm}
\begin{tabular}[h]{|c|c|l|}
\hline
Spectral value $E_n$ & $\int\limits_{0}^\infty \psi_n(t)^2~dt$ & $\left[\lim\limits_{x \rightarrow 0} W_{\psi,\frac{\partial}{\partial E} \psi}(x) -
\lim\limits_{x \rightarrow \infty} W_{\psi,\frac{\partial}{\partial E} \psi}(x) \right]_{\mid E=E_n}$ \\[0ex]
\hline
$E_0 ~=~-3.842367$ & $5.38183 \times 10^{-12}$ & $5.38183 \times 10^{-12}$ \\[1ex]
\hline
$E_1 ~=~-1.319311$ & $8.40188\times 10^{-9}$ &  $8.40188\times 10^{-9}$ \\[1ex]
\hline
$E_2 ~=~-0.505219$ & $2.46982\times 10^{-6}$ & $2.46982\times 10^{-6}$ \\[1ex]
\hline
$E_3 ~=~-0.161364$ & $0.00173697$ & $0.00173697$ \\[1ex]
\hline
$E_4 ~=~-0.024529 $ & $18.2922$ & $18.2922$ \\[1ex]
\hline
\end{tabular}
\caption{Evaluation of the integral formula (\ref{intpsi}) for the parameter setting $\sigma=-x_0=V_0=5$.} \label{table}
\end{center}
\end{table}
Under the setting taken in table \ref{table} the discrete spectrum of the boundary-value problem (\ref{bvp1}), (\ref{bvp2})
contains five values.
For each of those (left column), we calculate the normalization integral numerically (middle column), and
evaluate the right side of our identity (\ref{intpsi}) (right column). We see that the same numerical results are
obtained from each side of the latter identity, confirming its correctness.

\paragraph{Double integration.} We will now apply the double integral formula (\ref{int2}). Maintaining the same
choice for the function $u$ as in the previous paragraph, our formula reads
\begin{eqnarray}
\int\limits_{p}^x \int\limits_{p}^{x_1} \left[\frac{\psi(x_2)}{\psi(x_1)}\right]^2 dx_2~dx_1 &=&
\frac{{\displaystyle{\frac{\partial}{\partial E}~\psi(p)}}}{\psi(p)}-
\frac{{\displaystyle{\frac{\partial}{\partial E}~\psi(x) }}}{\psi(x)}+
W_{\psi,\frac{\partial}{\partial E}\psi}(p)~
\int\limits_{p}^x \frac{1}{\psi(t)^2}~dt. \label{intpsi2}
\end{eqnarray}
Recall that $\psi$ refers to the function from (\ref{solfull}), where $E$ is not yet replaced by a spectral value $E_n$.
As in case of (\ref{intpsi}) we assume that the integrals exist, which is not guaranteed a priori, but depends on the
choice of parameters that $\psi$ contains. An inconvenient feature of (\ref{intpsi2}) is the integral on the right side. We
can avoid its calculation by using the reduction-of-order formula: if $v$ is a solution of (\ref{bvp1}), such that
$\psi$ and $v$ are linearly independent and have a Wronskian equal to one, then
\begin{eqnarray}
\int\limits_p^x \frac{1}{\psi(t)^2}~dt &=& \frac{v(x)}{\psi(x)}-\frac{v(p)}{\psi(p)}. \label{red}
\end{eqnarray}
A possible choice for the function $v$ can be constructed using the results in \cite{artur}. There it was shown that
two linearly independent solutions $\psi$ and $\psi^\dagger$ of (\ref{bvp1}) are given by (\ref{solder}) and by the
same function (\ref{solder}), provided ${}_1F_1$ is replaced by Tricomi's confluent
hypergeometric function \cite{abram}, respectively. Now we set
\begin{eqnarray}
v(x) &=& \frac{\psi^\dagger(x)}{W_{\psi,\psi^\dagger}(x)}, \label{vli}
\end{eqnarray}
which yields a function that has the desired properties, in particular, the Wronskian of $\psi$ and $v$ equals one.
Substitution of (\ref{red}), (\ref{vli}) and evaluating the right side of (\ref{intpsi2}) yields an enormous expression that
we do not show here. We point out, however, that this expression is an integral-free representation of the
double integral in (\ref{intpsi2}). Table \ref{table2} shows a numerical evaluation of the latter double integral, as well as
a comparison with the corresponding right side of (\ref{intpsi2}).
\begin{table}[h]
\begin{center}
\vspace{.5cm}
\begin{tabular}[h]{|l|l|l|}
\hline
Parameter value $E$ & $\int\limits_{0.1}^1 \left[\frac{\psi(x_2)}{\psi(x_1)}\right]^2 dx_2~dx_1$ & Right side of
(\ref{intpsi2}) \\[0ex]
\hline
$-3.842367$ & $0.280879$ & $0.280879$ \\[1ex]
\hline
$-1.319311$ & $0.540471$ &  $0.540471$ \\[1ex]
\hline
$-0.505219$ & $0.838775$ & $0.838775$ \\[1ex]
\hline
$-0.161364$ & $1.11869$ & $1.11869$ \\[1ex]
\hline
$-0.024529 $ & $1.29632$ & $1.29632$ \\[1ex]
\hline
\end{tabular}
\caption{Evaluation of the integral formula (\ref{intpsi2}) for the parameter setting $\sigma=-x_0=V_0=5$.} \label{table2}
\end{center}
\end{table}
Note that in the left column we used the same spectral values as in our previous table \ref{table}.
Since this time the integrand on the left side of (\ref{intpsi2}) does not have any immediate physical meaning, we do not
have to pick spectral values, but can choose any value of $E$. The reason for reusing the spectral values is that for these
values the function $\psi$ does not have many zeros, such that the integral in the middle column of \ref{table2}
exists. Comparison of the middle and the right column confirms that our integral formula (\ref{intpsi2}) works
correctly. Before we conclude this section let us recall that the purpose of the latter formula is the construction of
indefinite integrals (definite integrals with undetermined integral limits), which due to their length we cannot
show here.

\section{The massless Dirac equation}
Our final generalization of Lambert-W systems is concerned with the massless Dirac equation in two dimensions.
We will show two different methods of converting a Dirac equation into its Schr\"odinger counterpart, thus
generating exactly-solvable systems for non-diagonal and diagonal potential matrices.

\subsection{Dirac equation with scalar potential: general solution}
The stationary massless Dirac equation in two dimensions can be written as
\begin{eqnarray}
\left[\sigma_1~p_x+\sigma_2~p_y + V(x)~I_2 \right] \Psi(x,y) &=& E~\Psi(x,y), \label{dirac0}
\end{eqnarray}
where $\sigma_j$, $j=1,2$, stand for the Pauli spin matrices and
$p_x, p_y$ represent the respective momentum operators. Furthermore, the potential $V$ is a smooth function,
$I_2$ represents the $2 \times 2$ identity matrix, the constant $E$ is the stationary energy of the system,
and $\Psi$ denotes the two-component solution. We can decouple our Dirac equation by setting
\begin{eqnarray}
\Psi(x,y) &=& \exp\left(i~k_y~y \right)~
\left[\Psi_1(x)+\Psi_2(x), \Psi_1(x)-\Psi_2(x) \right]^T \label{psinew0} \\[1ex]
\Psi_2(x) &=& -\frac{i}{k_y} \left\{
i~\Psi_1'(x)-\left[E-V(x) \right] \Psi_1(x)\right\}. \label{psi20}
\end{eqnarray}
Here the wave number $k_y$ describes free motion in the $y$-direction, and the function $\Psi_1$ is a solution
to the Klein-Gordon type equation
\begin{eqnarray}
\Psi_1''(x) + \left\{ \left[V(x)-E\right]^2-k_y^2+i~V'(x)\right\} \Psi_1(x) &=& 0. \label{sse0}
\end{eqnarray}
In the next step, let us incorporate the potential $V$ from our initial model (\ref{bvp1}), as given in (\ref{pot}).
Upon substitution of this potential, equation (\ref{sse0}) renders exactly-solvable \cite{tarloyan}. The
general solution can be expressed as follows
\begin{eqnarray}
\Psi_1(x) &=& z^\frac{\gamma}{2}~\exp\left(\frac{\delta}{2}~z\right)~\frac{d}{dz}
\left\{c_1~
\exp\left[-\frac{\delta+s_0}{2}~z \right] {}_1F_1\left(\alpha,\gamma,s_0~z \right)
+ \nonumber \right. \\[1ex]
& & \left. \hspace{3cm}
+~c_2~
\exp\left[-\frac{\delta+s_0}{2}~z \right] U\left(\alpha,\gamma,s_0~z \right)
\right\}_{\Big| z = W[-\exp(-\frac{x-x_0}{\sigma})]}. \label{psi1gen}
\end{eqnarray}
Note that the symbols ${}_1F_1$ and $U$ represent the Kummer and Tricomi confluent hypergeometric functions,
respectively \cite{abram}. Furthermore, $c_1, c_2$ stand for arbitrary linear factors, and the following
abbreviations are in use
\begin{eqnarray}
\begin{array}{llllllllll}
K_0 ~=~ \sqrt{k_y^2-(E-V_0)^2} &~~~ K_1 ~=~ \sqrt{k_y^2-E^2} &~~~
\alpha ~=~ \frac{\sigma}{2~K_0} \left[(K_0+K_1)^2+V_0^2 \right] \\[1ex]
\gamma ~=~ 2~\sigma~K_1 &~~~
\delta ~=~ 2~\sigma~(K_1+i~V_0) &~~~ s_0 ~=~ 2~\sigma~K_0. \nonumber
\end{array}
\end{eqnarray}
We can now obtain the general solution of our Dirac equation (\ref{dirac0}) for the potential (\ref{pot}) by
plugging the function $\Psi_1$ as given in (\ref{psi1gen}), into the two-component vector (\ref{psinew0}) and
the second component (\ref{psi20}).

\subsection{Dirac equation with matrix potential: bound states at zero energy}
\begin{figure}[h]
\begin{center}
\epsfig{file=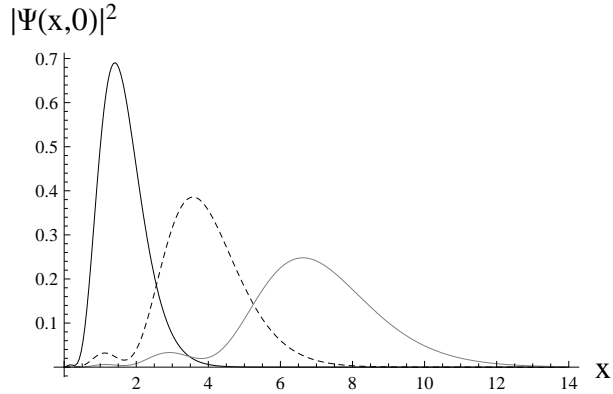,width=8cm}
\caption{Graphs of the normalized probability density associated with the solutions (\ref{psinew})-(\ref{psi2}) for
the parameter settings $\sigma=-x_0=V_0=5$, $V_{21}(x)=i/x$, $V_{22}(x)=1$, and $k_y = \sqrt{3.842367}$ (solid black curve),
$k_y = \sqrt{1.319311}$ (dashed curve), $k_y = \sqrt{0.505219}$ (gray curve).}
\label{figdiraczero}
\end{center}
\end{figure}

Let us now consider an example that involves a nondiagonal matrix potential.
As in the previous case we start out from the massless Dirac equation in the form
\begin{eqnarray}
\left[\sigma_1~p_x+\sigma_2~p_y + V(x) \right] \Psi(x,y) &=& 0, \label{dirac}
\end{eqnarray}
where we adopt the notation from (\ref{dirac0}). This time, however, the potential $V=(V_{ij})$ is a $2 \times 2$
matrix that does not depend on the energy $E$. We make the definitions
\begin{eqnarray}
\Psi(x,y) &=& \exp\left(i~k_y~y \right)~
\left[\Psi_1(x),\Psi_2(x) \right]^T \label{psinew} \\[1ex]
\Psi_1(x) &=& \exp\left[-\frac{1}{2}~ \int\limits^x i~V_{12}(t)+i~V_{21}(t)-\frac{V_{22}'(t)}{V_{22}(t)}~dt\right]
\psi(x) \label{psi1}\\[1ex]
\Psi_2(x) &=& \frac{i}{V_{22}(x)}~\Psi_1'(x)-\left[
\frac{i~k_y+V_{21}(x)}{V_{22}(x)}\right] \Psi_1(x), \label{psi2}
\end{eqnarray}
where the function $\psi$ solves the following equation
\begin{eqnarray}
& &\psi''(x)-\Bigg\{k_y^2+k_y \left[i~V_{12}(x)-V_{21}(x)-\frac{V_{22}'(x)}{V_{22}(x)}\right]
-\frac{V_{12}(x)^2+V_{21}(x)^2}{4}+\frac{V_{12}(x)~V_{21}(x)}{2}- \nonumber \\[1ex]
& & \hspace{1.1cm} -~V_{11}(x)~V_{22}(x)+
\frac{i~V_{12}'(x)-i~V_{21}'(x)}{2}+
\frac{i~V_{21}(x)~V_{22}'(x)-i~V_{12}(x)~V_{22}'(x)}{2~V_{22}(x)}+ \nonumber \\[1ex]
& & \hspace{1.1cm} +~\frac{3~V_{22}'(x)^2}{4~V_{22}(x)^2}
-\frac{V_{22}''(x)}{2~V_{22}(x)}
\Bigg\}~\psi(x) ~=~ 0. \label{sse}
\end{eqnarray}
Since this equation has Schr\"odinger form, we can match it with (\ref{bvp1}). This results in the
setting $E=-k_y^2$ and in the following constraints
\begin{eqnarray}
0 &=& i~V_{12}(x)-i~V_{21}(x)-\frac{V_{22}'(x)}{V_{22}(x)}  \label{sys1} \\[1ex]
V_0-\frac{V_0}{1+W \hspace{-.1cm}\left[
-\exp\left(
\frac{x_0-x}{\sigma}
\right)
\right]
} &=&
-\frac{V_{12}(x)^2+V_{21}(x)^2}{4}+\frac{V_{12}(x)~V_{21}(x)}{2} -~V_{11}(x)~V_{22}(x)+ \nonumber \\[1ex]
& & +~\frac{i~V_{12}'(x)-i~V_{21}'(x)}{2}+
\frac{i~V_{21}(x)~V_{22}'(x)-i~V_{12}(x)~V_{22}'(x)}{2~V_{22}(x)}+ \nonumber \\[1ex]
& & +~\frac{3~V_{22}'(x)^2}{4~V_{22}(x)^2}
-\frac{V_{22}''(x)}{2~V_{22}(x)}. \label{sys2}
\end{eqnarray}
We can solve these constraints for the components of the Dirac matrix potential. The solution of the
system (\ref{sys1}), (\ref{sys2}) can be written in the form
\begin{eqnarray}
V_{11}(x) &=& -\frac{1}{V_{22}(x)}
\left\{
V_0-\frac{V_0}{1+W \hspace{-.1cm}\left[
-\exp\left(
\frac{x_0-x}{\sigma}
\right)
\right]
}
\right\}
 \label{ssys1} \\[1ex]
V_{12}(x) &=& V_{21}(x) -i~\frac{V_{22}'(x)}{V_{22}(x)}. \label{ssys2}
\end{eqnarray}
We note that in this representation the two potential components $V_{21}$ and $V_{22}$ remain
undetermined and can be chosen arbitrarily. Let us now construct a solution of the Dirac equation (\ref{dirac}) that
pertains to the matrix potential $V$ specified in (\ref{ssys1}), (\ref{ssys2}). To this end, we note that the
function $\psi$ in (\ref{psi1}) can be chosen as the solution $\psi_n$ in (\ref{solfull}) to our boundary-value problem
(\ref{bvp1}), (\ref{bvp2}). The resulting expressions (\ref{psinew})-(\ref{psi2}) that define the Dirac solution are
very long, such that we do not show them here. Instead, in figure \ref{figdiraczero}
we display graphs of three normalized probability densities associated with our Dirac solution.
 \noindent
The parameters
are chosen such that we can obtain the corresponding values for $k_y$ from the spectral values $E_n$ given in
table 1 by means of the relation $E_n=-k_y^2$. In addition we chose the free entries $V_{21}$ and
$V_{22}$ of the Dirac matrix potential as $V_{21}(x)=i/x$, $V_{22}(x)=1$. Note that these settings render the
off-diagonal potential entries imaginary, which in the present example is necessary to generate bound-state solutions.

\section{Concluding remarks}
In this work we have constructed and analyzed various generalizations of a Schr\"odinger system
involving a singular Lambert-W potential. Our results apply to both Schr\"odinger and Dirac models,
complementing recent findings related to the Klein-Gordon equation \cite{tarloyan}. It should be
pointed out that our examples are far from being exhaustive in several ways. First, different point transformations
or higher-order SUSY transformation can be applied to generate additional exactly-solvable Schr\"odinger
models, which in turn give rise to exactly-solvable Dirac counterparts. Second, our methods can be applied to
systems different from the initial one (\ref{bvp1}), (\ref{bvp2}) considered here, for example the Lambert-W step model
\cite{artur2}. Third, particularly Dirac models can be generalized by directly performing Darboux transformations
without taking into account the associated Schr\"odinger systems. These topics will be subject to future research.

\section{Acknowledgments}
 A.M. Ishkhanyan acknowledges the support by the Russian-Armenian University and the Armenian Science Committee (SC Grants No. 18RF-139 and
 No. 18T-1C276).

\end{sloppypar}

\end{document}